\documentclass[12pt, preprint]{aastex}
\usepackage{graphicx}
\shorttitle{Double coronal X-ray sources of a magnetic breakout}
\shortauthors{Chen et al.}
\usepackage{color}
\usepackage{multirow}
\begin{document}

\title{Double Coronal X-ray and Microwave Sources Associated With A Magnetic Breakout Solar Eruption}

\author{Yao Chen\altaffilmark{1}, Zhao Wu\altaffilmark{1}, Wei Liu\altaffilmark{2}, Richard A. Schwartz\altaffilmark{3}, Di Zhao\altaffilmark{1}, Bing Wang\altaffilmark{1}, Guohui Du\altaffilmark{1}}

\altaffiltext{1}{Shandong Provincial Key Laboratory of Optical
Astronomy and Solar-Terrestrial Environment, and Institute of
Space Sciences, Shandong University, Weihai, Shandong 264209,
China; yaochen@sdu.edu.cn}

\altaffiltext{2}{W. W. Hansen
Experimental Physics Laboratory, Stanford University, Stanford, CA
94305, USA}

\altaffiltext{3}{NASA Goddard Space Flight Center and American
University, Greenbelt, MD 20771, USA}

\begin{abstract}
Double coronal hard X-ray (HXR) sources are believed to be
critical observational evidence of bi-directional energy release
through magnetic reconnection in a large-scale current sheet in solar flares. Here we present a study on double coronal sources observed in both HXR and microwave
regimes, revealing new characteristics distinct from earlier
reports. This event is associated with a footpoint-occulted
X1.3-class flare (25 April 2014, starting at 00:17 UT) and a coronal mass ejection that are likely
triggered by the magnetic breakout process, with the lower source
extending upward from the top of the partially-occulted flare loops and the upper source co-incident with rapidly squeezing-in side lobes (at a speed of $\sim$250 km s$^{-1}$ on both sides). The upper source can be identified at energies as high as 70-100 keV. The X-ray upper source is characterized by flux curves different from the lower source, a weak energy dependence of projected centroid
altitude above 20 keV, a shorter duration and a HXR photon spectrum slightly-harder than those of the lower source.
In addition, the microwave emission at 34 GHz also exhibits a similar double source structure and the
microwave spectra at both sources are in line with
gyro-synchrotron emission given by non-thermal energetic
electrons. {These observations, especially the co-incidence of the very-fast squeezing-in motion of side lobes and the upper source, indicate that the upper source is associated with (possibly caused by)
this fast motion of arcades. This sheds new lights on the origin of the corona double-source structure observed in both HXRs and microwaves.}
\end{abstract}


\section{Introduction}
During solar flares, a significant amount of magnetic energy is
released to accelerate electrons (and other particles) through
magnetic reconnection in the corona. The accelerated electrons
stream down and bombard the lower chromospheric atmosphere to emit
hard X-rays (HXRs), as imaged by instruments such as the
\emph{Ramaty High Energy Solar Spectroscopic Imager}
(\emph{RHESSI}; Lin et al. 2002), while the intrinsic acceleration
site in the corona remains much more difficult to detect due to
its relatively faint emissivity and the limited dynamic range of
the instrument (see the reviews by Krucker et al. 2008, Fletcher
et al. 2011, White et al. 2011). Despite these limitations, major
steps toward a full understanding of coronal sources have been
achieved.

Masuda et al. (1994) discovered with the Yohkoh satellite a HXR
coronal source at energy level of 33$-$53 keV, located above the
soft X-rays (SXRs) loop top source. This and followup studies
(e.g., Tomczak 2001; Petrosian et al. 2002) found that the coronal
X-ray sources at higher energies appear at higher altitudes. This
has been interpreted with the standard picture of solar flares
(e.g., Priest {\&} Forbes 2002), in which higher flare loops are
filled with newly reconnected and heated plasmas and energetic
electrons of harder spectra, while lower loops are cooler of
previously heated plasmas with energetic electrons of softer
spectra. The Masuda-type above-the-SXR-loop-top HXR source (or
simply referred to as the above-the-loop-top source) is regarded
as the extreme of this trend (see, e.g., Liu et al. 2013).

Sui {\&} Holman (2003) discovered a second HXR coronal source
lying above the loop-top source. The combination of the two
coronal sources, when present simultaneously, are referred to as
the double-source structure. They and a series of following
studies (e.g., Sui et al. 2004; Veronig et al. 2006; Liu et al.
2008; 2013; Chen {\&} Petrosian 2012) found some peculiar
characteristics of this structure, such as energy-dependent
centroid locations with the sources getting closer at higher
energies (or temperatures), similar temporal variations of HXR
fluxes, and close energy spectral indices of the two sources.
These observations have been taken as major evidence supporting
the existence of a reconnecting current sheet (lying in between the
two sources) that acts as the expected bidirectional energy release
site of solar flares.

It should be noted that clear observational examples of double
coronal X-ray sources are rare and usually require limb
occultation of bright footpoint sources, due to the relative
faintness of coronal sources and the limited dynamic range of
present HXR imaging instruments. These occulted events provide a
major source of our knowledge on coronal sources (see the review
on this topic by Krucker et al., 2008).

Solar flares are closely related to coronal mass ejections (CMEs). One leading theory of the CME
initiation is the so-called magnetic breakout (Antiochos et al.
1999, Antiochos 1998), through which the coronal confinement
restraining the eruption can be released via the breakout
reconnection occurring high in the corona, within a multi-polar
magnetic configuration. The breakout reconnection sets in between
two sets of constraining field lines, including the large-scale
overlying field lines of the coronal background and the lower
arcade of loops that is immediately above the energized core flux
system. The core flux consists of a set of sheared or twisted
field lines, representative of the magnetic driver or energy
source of the eruption. The breakout process can be separated into
four evolutionary stages according to previous numerical
simulations (e.g., Lynch et al. 2004, 2008, 2009; MacNeice et al.
2004; Devore {\&} Antiochos 2008; Karpen et al. 2012): the initial
energy storage process occurring within the central core flux, the
breakout reconnection process as just elaborated, the impulsive
eruption process during which the flare starts impulsively and the
core flux takes off and expands outward while pushing aside nearby
side lobes, and the restoration stage which involves the so-called
anti-breakout reconnection in the aftermath of the CME evacuation
(e.g., Lynch et al., 2008). The anti-breakout reconnection sets in
between side lobes to restore the pre-eruption magnetic breakout
configuration, after the central arcade is restored through the
primary flaring reconnection.

Observational tests of the breakout model have been published by
many authors (e.g., Aulanier et al. 2000; Gary {\&} Moore 2004;
Harra et al. 2005; Mandrini et al. 2006; Aurass et al 2011, 2013; Shen
et al. 2012; Sun et al. 2013). Most studies focus on disk events
so as to examine the magnetic configuration. Chen et al. (2016)
present a breakout event observed from a novel limb perspective (on April 25 2014,
the same event investigated here), revealing many
features analogous to simulations. In their study, the presence of
breakout reconnection is supported by the overall
remarkably-analogous-to-simulation configuration as viewed from
the hot passbands (131 and 94~\AA{}) of the Atmospheric Imaging
Assembly (AIA; Lemen et al. 2012) on board the \emph{Solar
Dynamics Observatory} (\emph{SDO}; Pesnell et al. 2012). New clues
are also found, such as the presence of heated and slowly-evolving
X-shaped morphology with pairs of cusp-like structures, the sequential brightening of loops around
this X feature, among other details.

Of particular interest here is the presence of double coronal HXR
sources during the impulsive stage of the event. The double
sources exhibit as the simultaneous presence of two coronal
sources, one lower and one upper, in X-ray images from
\emph{RHESSI}. The lower source is a typical one that is immediately above the
partially-occulted flaring loops while the upper source is co-incident with
the re-approaching side lobes upon the rapid evacuation of the ejecta. {Note
that the pre-eruption breakout reconnection is not accompanied by any considerable
enhancement of X-ray and microwave emission, indicating insufficient energetic
electrons accelerated during that earlier process. The upper X-ray source, of
major interest here, is observed during the impulsive stage above the site of
the earlier breakout reconnection.} Chen et al. (2016)
presented a preliminary analysis and found that the upper source
reaches an energy level of 70-100 keV. To further explore its nature,
in this study we examine the spatially-resolved X-ray imaging spectra,
the energy dependence of the source centroid separation,
flux variation, as well as the simultaneous microwave sources recorded by the Nobeyama
Radioheliogram (NoRH; Nakajima et al. 1994; Takano et al. 1997) at 17 and 34~GHz.

\section{Event Overview}\label{sec2}
The complete evolutionary process of the breakout event has been
described in Chen et al. (2016). The event was originated from NOAA AR
12035 on the backside of the Sun (the southwestern quadrant) with the
accompanying flare starting at 00:17 UT on 25 April 2014. The
eruption is observed by the \emph{Solar TErrestrial RElations
Observatory} (\emph{STEREO}; Kaiser et al. 2008) B also as a
backside event and STEREO-A as a disk event on the southeastern
quadrant, with a 5-minute cadence at 195~\AA{} and a 10-minute cadence at 304~\AA{}.
According to STEREO-A, the center of the flare ribbons
is located at S20E60, $\sim$7$^\circ$ behind the limb from the
SDO-Earth perspective, corresponding to an occultation height of
$\sim$5 Mm (7''). Thus, a significant part of the flare emission
is occulted by the limb. This is why we can observe the faint
coronal emission with RHESSI in spite of its limited dynamic range.
{Note that the STEREO images are largely saturated
with large bright regions over the AR during the impulsive stage
of the X-1.3 flare of study. Thus, although these data are
valuable at inferring the overall morphology of the eruption from different perspectives,
like the occultation height as described above, they are
not helpful to our study on the origin of HXR emission
that lasts only a few minutes.}

In Figure 1, we show the AIA images in its hot passbands of 131
and 94~\AA{} during the impulsive stage. According to Chen et al.
(2016), the pre-impulsive stage of the event is characterized by
an at-least 40-minute-long breakout reconnection process
{occurring across an X-shape structure at a height of 70'' above the solar limb}. The
subsequent outward eruption of the core flux (black arrow)
transforms the reconnecting structure into an extended-curved
transverse structure of a high temperature (yellow arrows),
presumably a current sheet. During this impulsive stage, the two
side lobes (red arrows) are pushed aside. Around 00:20 UT when the
bright CME core rises to a height of 50'', the side lobes start to
move towards each other. This forms a spearhead-like morphology in
the distance-time maps along the three slices (S1-S3, see Figure
1c) that capture the dynamics of the lobe-approaching region.
The speeds of the loop at the upper side along the slices vary from 220-270 km s$^{-1}$, comparable with those measured at the lower side. This gives an approaching (or squeezing-in) speed of loops about twice the above value.

Note that approaching coronal structures during a solar flare are frequently observed. Measurements of their speeds have been used to infer the reconnection inflow speed. For example, Lin et al. (2005) deduced that the average reconnection inflow velocities near the presumed current sheet over different time intervals ranged from 10.5 to 106 km s$^{-1}$, \textbf{Hara et al. (2011) deduced a flow into the presumed reconnection region (the loop top) with a Doppler velocity of 20 km s$^{-1}$, }and Su et al. (2013) obtained apparent inflow velocities of $\sim$50 km s$^{-1}$ from the south and $\sim$20 km s$^{-1}$ from the north side of the reconnection site. These values are considerably less than those measured here. This rapid squeezing-in motion even results in an inverted-pear like shape of
the rear of the ejecta and a second clear X-shaped structure high in the
corona ($\sim$40''$-$50'' above the limb). {This newly-formed X-shaped structure,
considerably lower than the earlier X structure of
the pre-flare triggering breakout reconnection, is coincident and co-spatial with
the upper HXR coronal source (the focus of this study).}

\section{HXR and microwave data analysis on the breakout double-source event}\label{sec3}
The new X-shape morphology and rapid squeezing-in motion are observed
during the impulsive stage, lasting for $\sim$2 minutes. After 00:22 UT, both features disappear from
the AIA images. During this 2-minute interval, RHESSI recorded a HXR
double-source structure. Before examining the spatially-resolved
HXR images, we first check the integrated light curves at
different energy bands as shown in Figure 2. It can be seen that
the 6-12 and 12-25 keV curves rise gradually from the start of the
plot (00:18 UT), basically following the rising trend of the two
GOES SXR curves, while inconsistent with their derivatives. This
suggests that these low-energy fluxes are dominated by thermal
emission. On the other hand, the high energy curves (25-50 and
50-100 keV) present impulsive increases just before 00:20 UT,
neither following the GOES curves nor their derivatives. This
fact, being inconsistent with the well-known Neupert effect
(Neupert 1968), may be attributed to the occultation of the footpoint
HXR emissions that are the cause of heating the SXR-emitting flare
loops, while here the coronal HXR sources have no contribution to
the earlier rising SXR emission. Note that the above phenomenon is
also reported by Effenberger et al. (2016) in their independent
study of the same event.

In Figure 1a, we overplotted the flux contours of the
spatially-resolved HXR data. We see that, as briefly discussed by
Chen et al. (2016), the upper source reaches an energy
level of 70-100 keV, {and the lower sources concentrate
around 10'' above the limb, while the upper sources are around 50''.
The two sources are spatially separated by $\sim$40''. The
lower source extends upward from the flare loop
top, and the upper source is co-spatial with the above-described X morphology. As seen from Figure 3, before 00:21:24 UT, the centroid distances of the lower source increase in general with increasing energy (below 50 keV). At higher energy (50-100 keV), the lower source (00:20:12 $-$ 00:21:16 UT) descends considerably. For the upper source, the energy dependence of the source centroids is not significant below 20 keV, while above 20 keV the upper source presents a weak first-rising-then-descending trend with increasing energy. Note that these distance measurements suffer from projection effects. The inferred energy dependence may be different if observed
from a different perspective.}

In Figure 4a-4b we plot the light curves within energy bands of
12-25, 25-50, and 50-100 keV for both sources. The areas used for the
plots are drawn in Figure 5a (see circles 1 and 2, and the circles
3 and 4 represent the areas used to calculate the background X-ray fluxes). We see that the lower-source 12-25 keV curve start to increase around 00:19 UT,
nearly 1 minute earlier than those of the upper source, the 25-50 and 50-100
keV curves of both sources rise almost simultaneously. Two minutes
later the upper source declines rapidly while the lower source
continues to be at a relatively high level. This indicates that
the upper source is more transient. We highlight that
the presence of the upper source at 25-50 and 50-100 keV bands is
co-temporal and co-spatial with the side-lobe squeezing-in
motion observed from 00:20-00:22 UT (see Figure 1c).

In the centroid distance plot of Figure 3, we integrate the 1-minute
RHESSI data to get enough photon counts within relatively fine
energy bands. To investigate the temporal behavior of the
double sources, we consider a shorter interval (20 s) of
integration yet using wider energy bands as a tradeoff. The
results are shown in Figure 4c. We see that the lower sources are
around 5-15'' at 00:21 UT with the 50-100 keV source being higher
untill 00:22 UT. After 00:22 UT, the lower sources manifest a slow
rising trend reaching $\sim$10-20'' at 00:25 UT. For the upper
source the centroid altitudes at higher energy tend to be slightly
higher initially, at the following two intervals the centroid altitudes
present a converging trend at $\sim$50'' above the solar limb.

Note that we tried three methods, including the PIXON (Metcalf et al. 1996), CLEAN
(Hurford et al. 2002), and VIS\underline{ }FWDFIT (Forward-fit algorithm based on
visibilities) algorithms (available in the RHESSI software) to reconstruct RHESSI images, and
found consistent results (see Aschwanden et al. (2004) for a comparative study with different algorithms). For the above source centroid measurements, we mainly use PIXON,
while doing imaging spectroscopy we use the three methods for a
consistency check.

As seen from Figure 5, the obtained three sets of spectral fits, each consisting of a thermal and a broken power-law component (the spectral index of the pre-break lower-energy part is taken to
be $-2$), present a spectral index (post-break) of the upper source slightly
smaller by $\sim$0.5 than that of the lower source. This hardening trend can
be clearly seen from the reconstructed photon spectral data from various methods.
Both sources are power-law dominated above $\sim$15 keV. The residuals (blue for the upper and red for the lower source) fluctuate around zero with an amplitude of $\sim$2$\sigma$. The CLEAN method yields in-general smaller residuals ($<1\sigma$ above 10 keV) than
the other two methods. Since we are mainly concerned
with the relatively harder spectra of the upper source
above 20 keV, the residuals of the spectral fits are in an acceptable range.

The event is also observed by the NoRH at 17 and 34~GHz with
synthesized images and the Nobeyama RadioPolarimeter (NoRP; Torii
et al. 1979) from 1-35 GHz with integrated flux densities. The
$T_B$ contours at the two frequencies have been overplotted onto
the 94~\AA{} images in Figure 1b, from which we see that the
34~GHz contours present a well-defined double source structure
that is largely coincident with the double HXR sources, while the
17~GHz data show an elongated feature possibly due to the coarser
angular resolution ($\sim$10'' at 17~GHz versus $\sim$5'' at
35~GHz).

The evolutionary sequence of NoRH sources is shown in Figure 6
with an accompanying movie. Note that the NoRH microwave data
presented here have a much higher temporal resolution (1s) than
the HXR imaging. This allows more source information to be
revealed. Initially at $\sim$00:20 UT, only weak emissions ($T_B
\leq 10^5$ K) are observed. The 17~GHz loop-top source increases its
$T_B$ rapidly to a level of $10^7$ K within 40s. Then, the 34~GHz
loop-top source also increases to the same level of $T_B$, while in the upper source
region the $T_B$s at both frequencies are much lower ($<10^6$ K). In general, the
17~GHz source is brighter than its 34~GHz counterpart. After
00:20:30 UT, the 34~GHz loop-top source starts to present an
extended structure. $\sim$10~s later, an isolated source appears
above the 34~GHz loop-top source, most evident around 00:21 UT.
The isolated source fades away after 00:21:30 UT.
Later, both 34-17~GHz sources maintain a bright localized loop-top
source. Through this process, the 17~GHz loop-top source shows
corresponding elongation towards the 34~GHz upper source, yet
without an isolated upper source structure.

In Figure 4d, we have plotted the $T_B$ profiles averaged within
the areas selected for the 17-34~GHz sources (see Figure 6, white curves). The
microwave profiles are consistent with their HXR
counterparts (25-50 and 50-100 keV), especially during the rising
phase. After the impulsive rise, the lower sources present a
double-hump feature. This is different from the upper source that
declines rapidly to the pre-eruption level in $\sim$2 minutes.

To determine the nature of the microwave emission and its
connection to the HXR emission, we examine the deduced microwave
spectra. The spatially-unresolved NoRP data can be used to measure
the peak (or turn-over) frequency of the total spectrum (see Figure
7a), which is important to determine the optical thickness of NoRH
sources. We see that the turn-over frequency during this period is
always $<$ 10 GHz. This turn-over frequency is mainly relevant to
the loop top source due to its dominance of emission. Considering
that the upper source is located much higher in the corona with a
general weaker magnetic field strength, it is
expected that the turn-over frequency of the upper source is even
lower (Dulk, 1985). This indicates that the 17-34~GHz data reported
here are optically-thin.

The NoRH $T_B$ spectral index ($\alpha$) map can be deduced with the measured $T_b$s at 17-34~GHz (see Figure 6 and the online movie). Values of $\alpha$ along the line S4 (shown in Figure 6c) are presented in Figure 7b. We see that during the presence of the upper source, $\alpha$ decreases to a value ranging from $-$3 to $-$4, a typical index for optically-thin gyro-synchrotron emission (e.g., Dulk et al. 1985; White et al. 2011; Narukage et al. 2014), from $\sim$2 for the lower source and from $\sim$0 for the upper source. This indicates that emissions from both sources include a significant non-thermal contribution.

The upper HXR source reported here is likely attributed to
thin-target bremsstrahlung emission since coronal plasmas are much
less dense than chromospheric plasmas of the thick-target footpoint
sources (see Krucker et al., 2008, 2010, 2014 for relevant
discussions). We then infer the electron energy spectral index
($\delta$) using the thin-target approximation with $\delta =
\gamma - 0.5$ (Brown 1971; Lin 1974; White et al. 2011), and get
$\delta$ = 3.6-3.8 ($\sim$3.1-3.3) for the lower (upper) source with the three methods.
On the other hand, with the non-thermal optically-thin gyro-synchrotron mechanism for the
microwave upper source with the $T_B$ spectral index $\alpha \sim -3.5$, we deduce the
electron energy spectra to be $\sim$3 using the Dulk
approximation (Dulk 1985). The
obtained HXR and microwave energy spectral indices of electrons
for the upper source are close to each other indicating
that both the HXR and microwave upper sources are generated by the
same population of energetic electrons. Note that the above analysis of
the electron energy spectra is preliminary and model-dependent.

The above discussion on HXR and microwave emission mechanisms helps
in understanding why the 34 GHz upper source has a brightness temperature ($\sim10^5 - 10^6$ K)
much lower than that of the lower source, and why the 34 GHz appears to be
displaced (by $\sim10''$) away from the source centroid of the counterpart
HXR source (see Figure 6). These observations may be due to the very different
dependence of HXR and microwave radiation on coronal parameters. The bremstrahlung
HXR emissivity is highly dependent on the electron number density while the
gyro-synchrotron microwave emission is proportional to a high power of
(thus very sensitive to the change of) the magnetic field strength (e.g., Dulk 1985).

To summarize this section, we find that (1) the double sources,
observed in both the HXR and microwave (34~GHz) regimes, are
co-incident with the rapid re-approaching motion of side lobes and
the presence of the resultant X-shape morphology; (2) the projected centroid
distances of the upper source present a weak energy dependence
above 20 keV; (3) the upper source is more transient lasting
for 1-2 minutes while the lower source is much longer in duration,
and their flux curves follow the almost-simultaneous rising trend
while the upper-source curve declines much sooner; (4) the HXR
upper source presents a spectrum with the power-law spectral index being slightly harder
(by $\sim$0.5) than that of the lower source, while in microwaves both sources present a typical non-thermal gyro-synchrotron emission with $T_b$ spectral
index ranging from $-3$ to $-4$.

\section{Summary and discussion}\label{sec5}
We investigated the flux and spectral characteristics of a coronal
double source structure observed in X-rays and microwaves during a foot-point occulted X-1.3 flare.
The event has been regarded as a clear example of magnetic
breakout observed from a limb perspective, during which the breakout reconnection high in the
corona removes the magnetic confinement of overlying arcades on
the lower energized central core flux and paves the way for the
eruption. The double sources, observed in HXRs by RHESSI and at
34~GHz by NoRH, appear during the impulsive stage of the event
when the ejecta moves outward to a certain height and the side
lobes that were previously pushed aside now start to move towards each other (or squeeze-in)
at a very fast pace ($\sim$220-270 km s$^{-1}$ on one side). We suggest that
this fast approaching motion of magnetic loops plays an important role
in generating the upper HXR and microwave coronal source.

%\textbf{(Remove these text: The side lobes are bright in the AIA 131 and 94~\AA{} while invisible
%in the cooler 171~\AA{} indicating that they carry high-temperature plasmas ($\sim6 - 10$ MK).
%These plasmas are likely heated through the earlier breakout reconnection.)}

The approaching motion of side lobes can effectively squeeze the field lines on the
two sides of the tail of the ejecta along which the presumed vertical current
sheet is located. This results in ``additional'' inward pressure that may have significantly
enhanced the rate of reconnection (driven by the very-fast inflows) and thus the electron acceleration at the wake of the ejecta. In addition, pre-accelerated electrons
(released by the on-going flare process) may be present as seed particles available for further
acceleration around the upper source. This presents a fast squeezing-in picture for the origin of the localised upper coronal source, and explains why the upper source is $\sim40''$ away from the lower source (among events with the largest double-source separation distance ever reported) with a power-law spectral index harder (by $\sim$0.5) than that of the lower source.

{In this scenario, the upper source is associated with the squeezing-in motion
of the side arcades. It is not necessarily located at the
upper tip of the reconnecting current sheet and related to the bidirectional energy release of reconnection across the large-scale current sheet, as usually assumed (e.g., Sui {\&} Holman 2003; Sui et al., 2004). In addition,
earlier studies found puzzling that the upper coronal source
can remain stationary in location for a few minutes before it
vanishes or moves outwards. Sui et al. (2004) tentatively attributed
this observation to the formation and development of the current sheet during the
early impulsive stage of the eruption, yet no further support was provided.
In our scenario, the upper source depends on the rapid approaching motion of
side lobes. The motion occurs within a relatively localized area, so does the
resultant upper source.

Note that the essential element of the above scenario,
i.e., the squeezing-in motion of arcades along the tail of the
ejecta, may take place as long as the eruption originates
from a multi-polar configuration, where surrounding arcades
are first pushed aside by the ejecta and then bounce backwards rapidly.
This may result in enhanced pressure towards the tail of the ejecta (likely
with a developing current sheet) and efficient reconnection and
electron acceleration therein, regardless of the exact
triggering process of the eruption (being a magnetic breakout or not).
Thus, the proposed scenario for the upper source is mainly
related to the global magnetic configuration rather than the
exact triggering mechanism. Further studies are required to see
whether more events are in line with this scenario.

%{\textbf{(Remove these text: According to numerical simulations of magnetic breakout (Lynch et al., %2008), during the flare impulsive stage approaching side lobes can get reconnected to restore the %pre-eruption magnetic configuration. This reconnection is named as the anti-breakout reconnection. %Thus, our study provides observational evidence for the occurrence of anti-breakout reconnection.)
%}

According to numerical simulations of magnetic breakout (Lynch et al., 2008), anti-breakout reconnection of side lobes sets in after central arcade is completely recovered through the
primary flaring reconnection. From the AIA data, during the impulsive stage (from 00:20-00:22 UT) coronal structures between the approaching side lobes brighten diffusively (see the online move provided in Chen et al. 2016). This makes it not possible to separate the anti-breakout reconnection from the primary flaring reconnection with the observations available here. We tentatively suggest that both may have contributed to the generation of the upper HXR source.

For the pre-eruption breakout reconnection (i.e., the triggering process), we did not observe any X-ray and radio signatures of efficient particle
acceleration. This is consistent with the deduction
that the triggering breakout reconnection mainly converts magnetic energy into
plasma thermal energy (Chen et al. 2016). During the impulsive stage, the pre-flare X
structure across which the triggering breakout reconnection takes place
deforms into a large-scale transversely-curved thin structure with a high temperature
(see the yellow arrows in Figure 1).
The structure has been suggested to be a transverse current sheet straddling the ejecta (see Chen et al. 2016). During its outward expansion, the structure keeps its shape and does not manifest
some usual reconnection-induced behavior, such as the presence of
magnetic islands, strong bidirectional flows, or some kind of morphology collapse, nor are there
any co-spatial X-ray or microwave sources. This indicates that the structure does not
experience fast reconnection and particles there are not efficiently accelerated.}

Aurass et al. (2013) presented an observational report on metric radio evidence of breakout reconnection during the impulsive stage of the flare occurring on 2003 November 3, on the basis of X-ray and radio-metric imaging and spectral data. Although their study was conducted in the SOHO era with less dynamical details and temperature coverage, the two studies together show that various reconnection processes associated with the breakout eruption can produce radio signatures at both microwave (at higher frequency and thus lower altitude) and metric (at lower frequency and thus higher altitude) wavelengths.

{In an independent study, Effenberger et al. (2016) investigated HXR emission of
the same event using the RHESSI data. They also performed an imaging spectroscopy
study of the double HXR sources and found that the upper source has a slightly
harder power-law spectral index. This allows them to speculate that there
exists effective electron acceleration in that part
of the loop (co-spatial with the upper source). This is consistent with
our result although the two studies have selected different time intervals of RHESSI
data, different regions for the imaging spectroscopy, and different fitting methods.}

{Double coronal sources associated with approaching loops were reported earlier
during the impulsive phase of a C2.3 confined flare (Su et al. 2013). In their event, clear
reconnection signatures of inflowing cool loops (at a speed of 20 (50) km s$^{-1}$ on the south (north) side)
and newly formed, out-flowing hot loops are observed. The X-ray energy of those coronal sources hardly reaches up to 20 keV, and the
X-ray spectra can be well fitted with two isothermal components. This indicates that
their coronal sources are likely thermal in nature. Here, our event is
associated with a major X-class flare and a fast CME, the X-ray energy in both coronal sources reach
up to 70 - 100 keV indicating the presence of nonthermal energetic electrons. The upper source
studied here is also co-incident in time and location with approaching loops, yet the process is
in the aftermath of a successful eruption, and the approaching speed is $\sim$5-10 times faster than those reported by Su et al. (2003). In addition, the side lobes of our event consist of high-temperature plasmas as inferred from the AIA data. This, again, presents our fast squeezing-in picture (driven by rapidly-approaching side lobes, with enhanced reconnection rate) for the origin of the upper coronal source.}

\acknowledgements

We are grateful to Dr. Brian Dennis for his constructive comments on the manuscript and to Ms. Kim Tolbert for her kind help with \emph{RHESSI} data analysis. We thank the \emph{SDO}, \emph{RHESSI} and NoRH-NoRP teams for the high-quality
EUV, X-ray, and microwave data. SDO is a mission of NASA¡¯s Living With a Star Program. This work was supported by the grant NNSFC 41331068.

\begin{figure}
\epsscale{1.}
\includegraphics[width=1.0\textwidth]{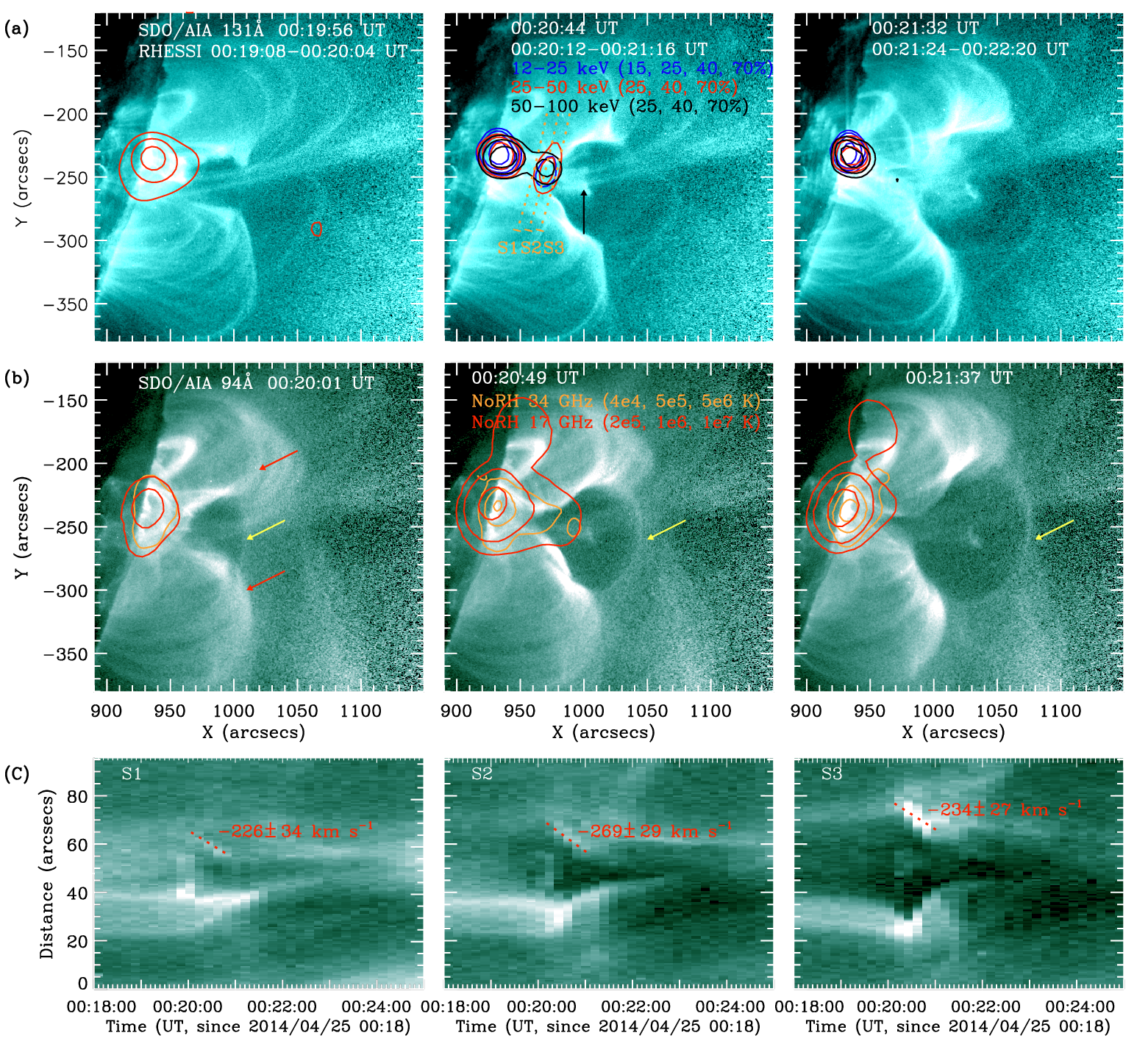}
\caption{Overview of the impulsive stage of the breakout event.
(a-b) AIA 131 and 94~\AA{} images, overlaid by (a) RHESSI contours
and (b) the 17-34~GHz microwave $T_B$ contours, with levels given
by percentages of the maximum HXR flux for corresponding energy band and the
microwave $T_B$ maxima. (c) distance-time maps for slices S1-S3
(the starting point is marked by a short dash). Dashed lines in
(c) represent linear fits to estimate velocities, with errors
given by a $\sim$4 pixel (1.74 Mm) uncertainty of distance
measurement. RHESSI images are reconstructed using the PIXON
algorithm and detectors 3, 5, 7-9. The black arrow in (a) points at
the bright ejecta, the yellow and red arrows in (b) point at the expanding
current sheet and the pair of side lobes, respectively.}\label{Fig1}
\end{figure}

\begin{figure}
\epsscale{1.}
\includegraphics[width=1.0\textwidth]{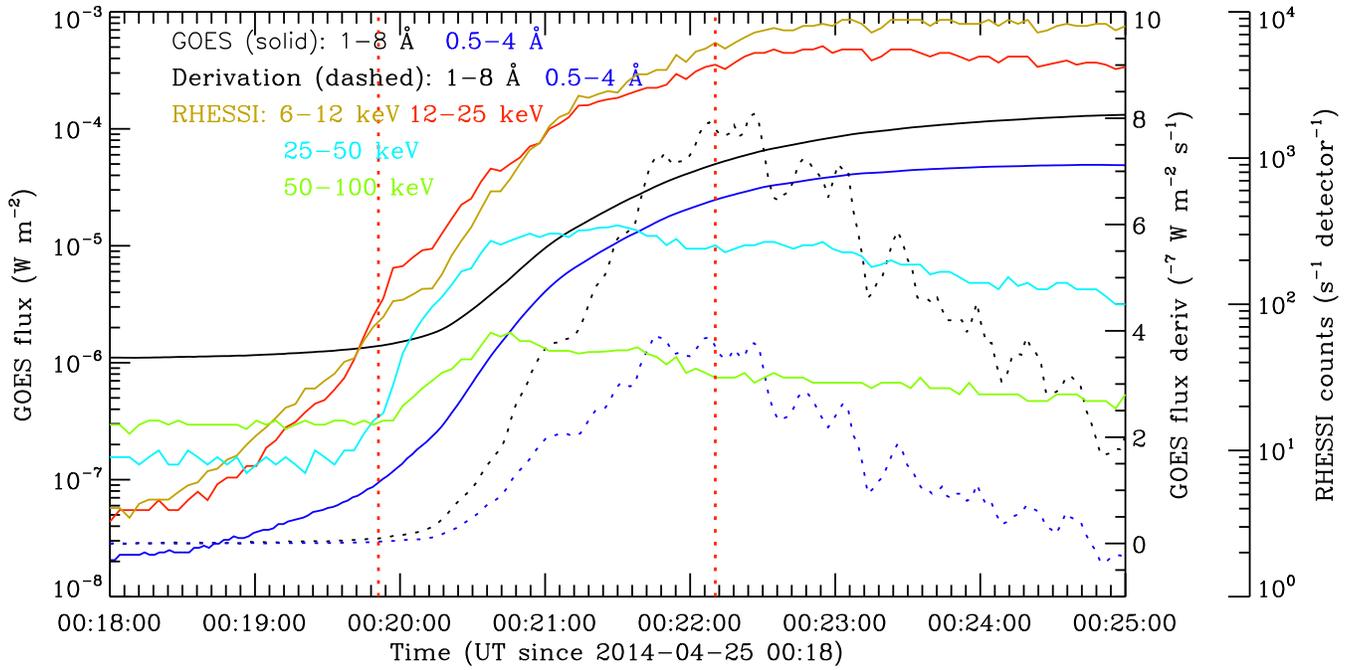}
\caption{GOES and RHESSI X-ray light curves (solid) and the GOES time derivatives (dashed) from 00:18 to 00:25 UT.}
\label{Fig2}
\end{figure}

\begin{figure}
\epsscale{1.}
\includegraphics[width=1.0\textwidth]{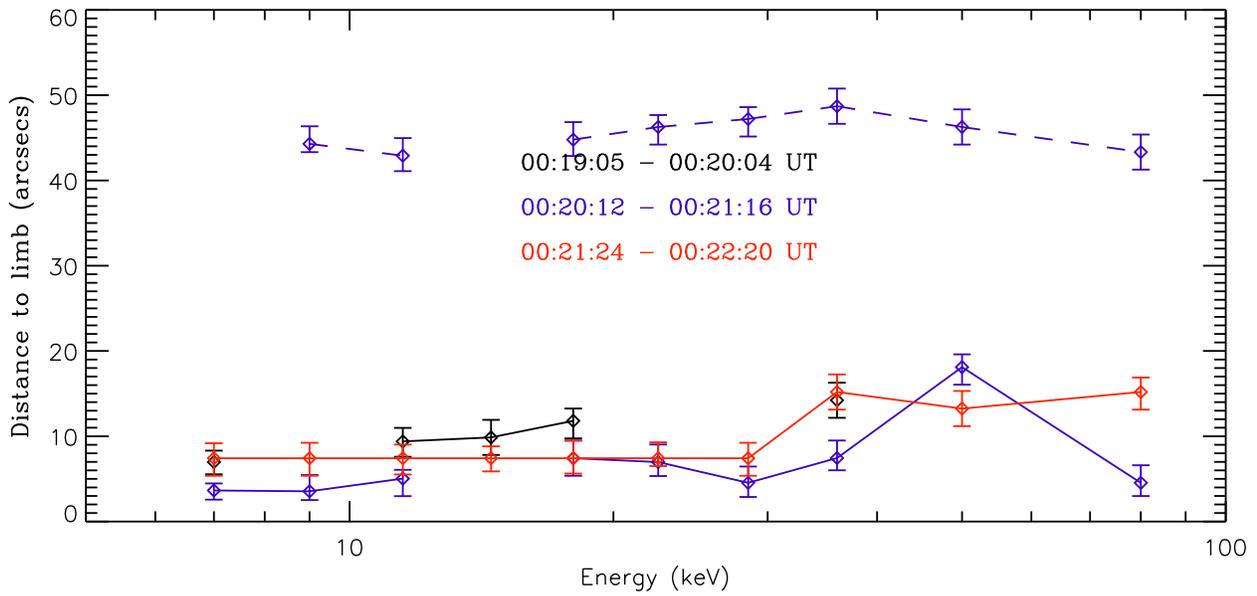}
\caption{Energy dependence of centroid distances of the upper and lower X-ray sources at different intervals. Error bars are given by the distance range of the 80{\%} level of the intensity maxima.}
\label{Fig3}
\end{figure}
%Error bars are obtained from the centroid position uncertainties
%in the same images reconstructed with the VIS\underline{ }FWDFIT (Forward-fit algorithm
%based on visibilities) algorithm currently available in the RHESSI software.

\begin{figure}
\epsscale{1.}
\includegraphics[width=1.0\textwidth]{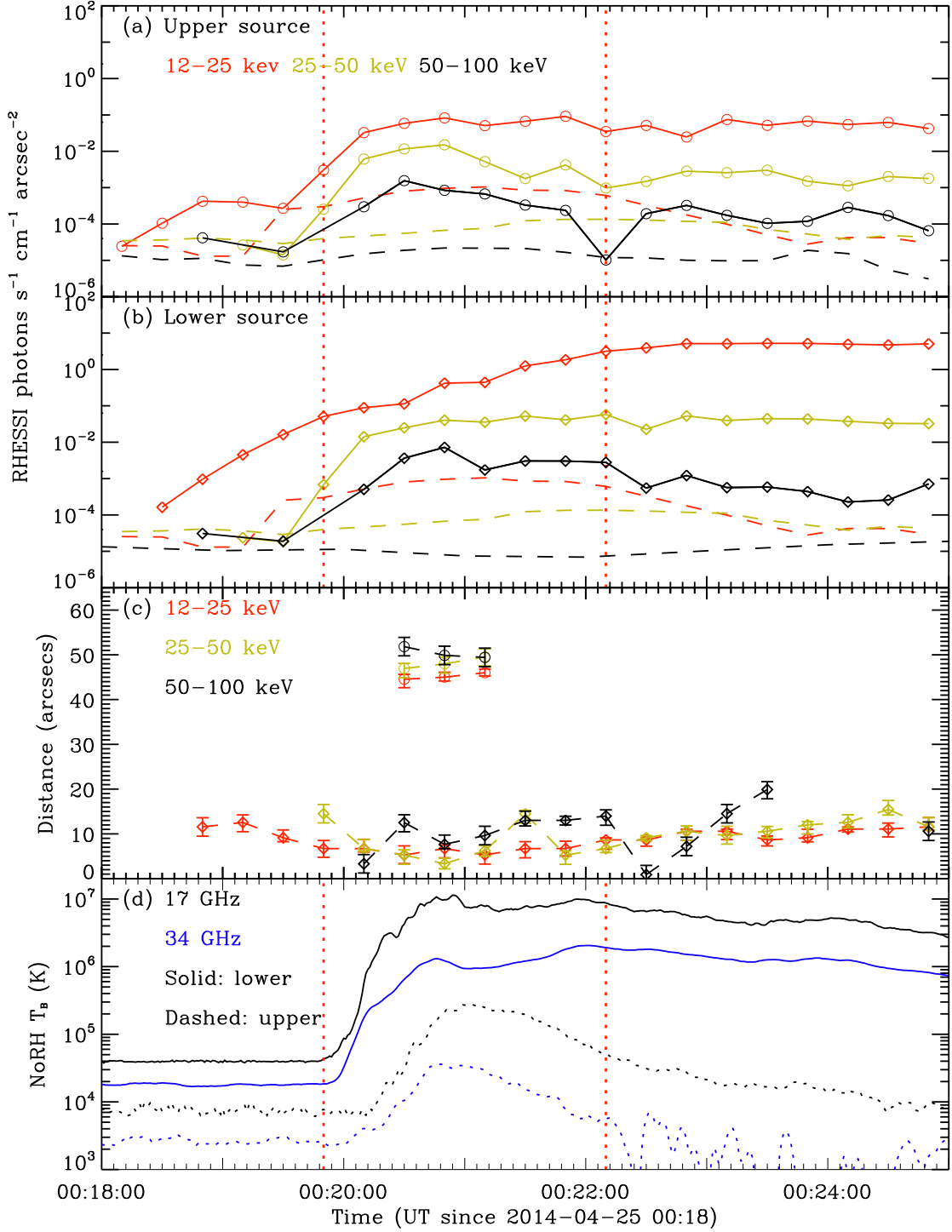}
\caption{RHESSI photon counts integrated over the circular areas plotted in Figure 5a for the upper (a) and lower (b) sources. The long dashed lines indicate the simultaneous HXR background levels, integrated over the circular areas (3 (lower) and 4 (upper)) as shown in Figure 5a. (c) The projected altitudes of source centroid, error bars are given by the distance range of the 80{\%} level of the intensity maxima. (d) NoRH $T_B$ maxima within the selected areas (white curves in Figure 6a). The two red vertical lines indicate the start and end of the 50-100 keV upper source.}\label{Fig4}
\end{figure}
%error bars are obtained from the centroid position uncertainties
%in the same images reconstructed with the VIS\underline{ }FWDFIT algorithm.

\begin{figure}
\epsscale{1.}
\includegraphics[width=.8\textwidth]{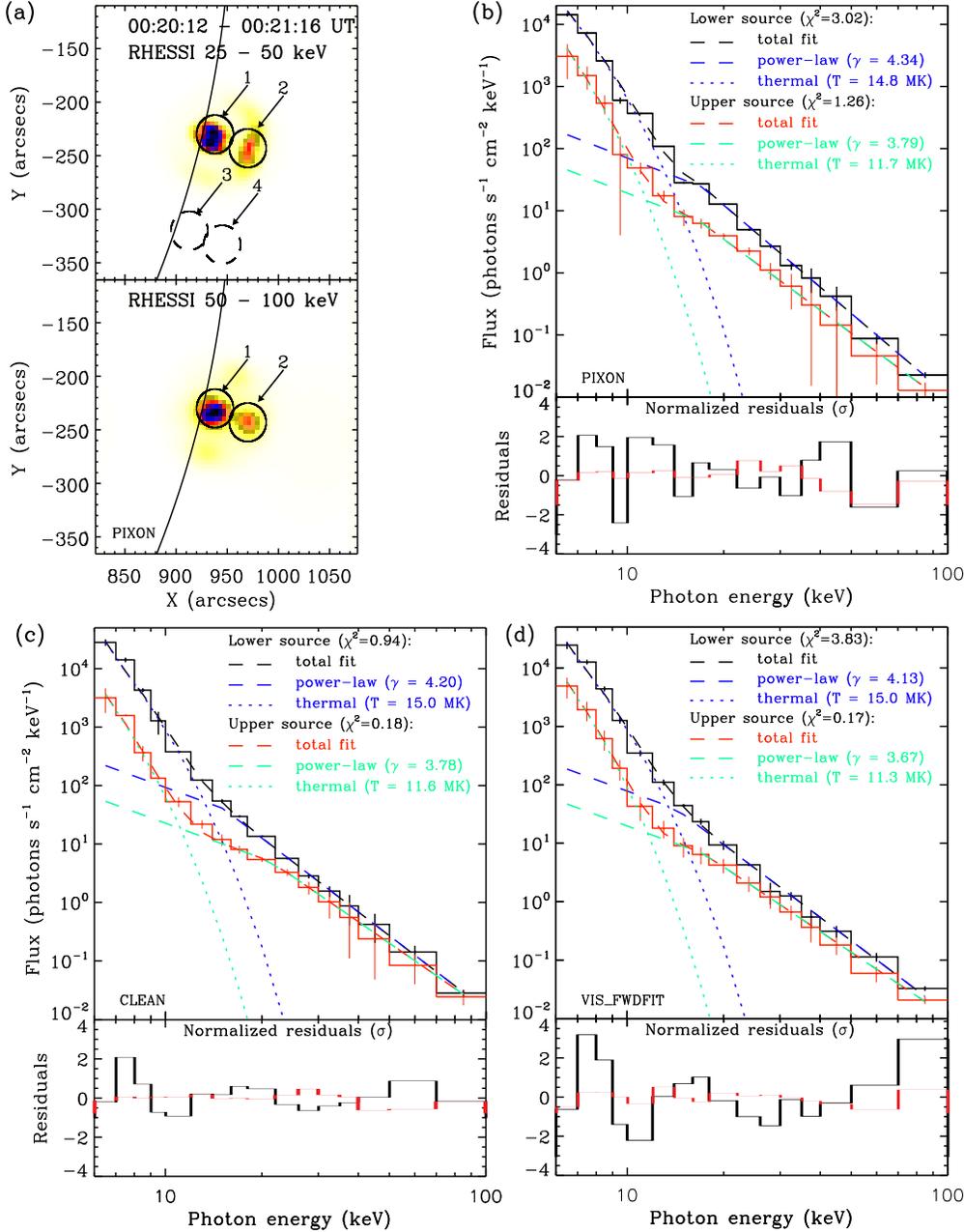}
\caption{(a) Hard X-ray images given by the PIXON method to indicate the selected
areas used to obtain the X-ray fluxes and imaging spectroscopy for the upper-lower (2-1) sources, and the corresponding background fluxes (4-3). (b-d) X-ray spectra and spectral fits of the double sources deduced with the three algorithms within selected areas. The spectral fits are given by a thermal (dotted) plus a broken power-law (dashed) model. The spectral index of the lower portion of the  power-law component is taken to be $-2$. Error bars (for the PIXON- and CLEAN- processed data) are given by the integration of data uncertainties of each pixel in the region of interest. For PIXON, the 2-D error map is determined with the RHESSI software (his\underline{ }calc\underline{ }image\underline{ }error.pro); for CLEAN, the RMS of the residual map is used to represent the uncertainty. For the VIS\underline{ }FWDFIT method, the error bars are given by {1/3} of the maximal fluxes outside the selected source areas. Residuals of the fit to the spatially integrated spectra, normalized to the 1$\sigma$ uncertainty of the measured flux at each energy band, are shown.}
\label{Fig5}
\end{figure}

\begin{figure}
\epsscale{1.}
\includegraphics[width=1.0\textwidth]{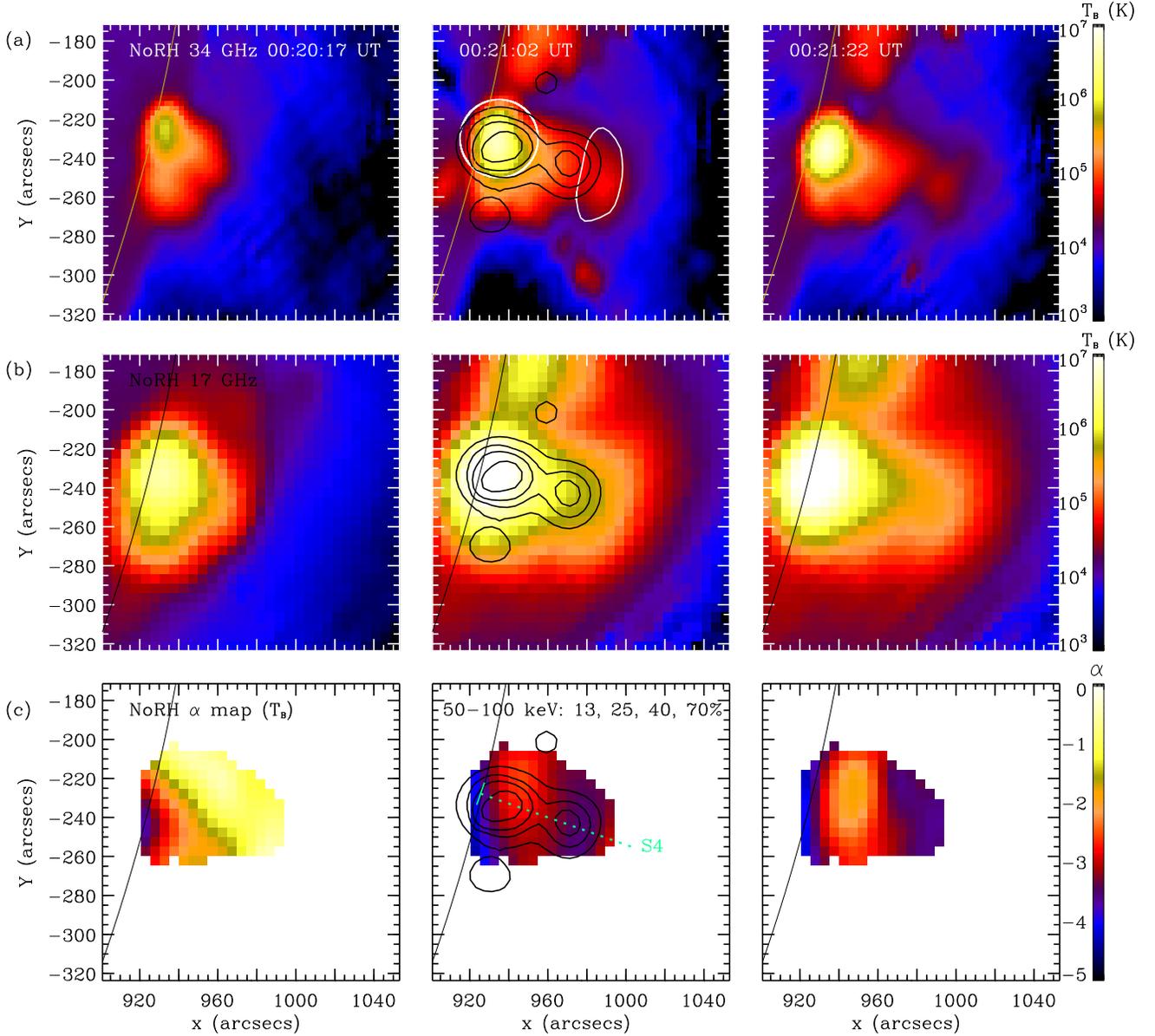}
\caption{The NoRH $T_B$ images and the deduced $T_B$
spectral indices ($\alpha$) at 00:20:17, 00:21:02, and 00:21:22 UT. The areas
given by the white curves are for the $T_B$ profiles presented in
Figure 4d, the slice S4 is for the $\alpha$ plots in Figure 7b. The
50-100 keV RHESSI contours are over-plotted in the middle panels
for comparison. The 12s raw NoRH data are integrated before
applying the CLEAN algorithm to reduce the system noise. (An
animation of this figure is available.) }\label{Fig6}
\end{figure}

\begin{figure}
\epsscale{1.}
\includegraphics[width=1.0\textwidth]{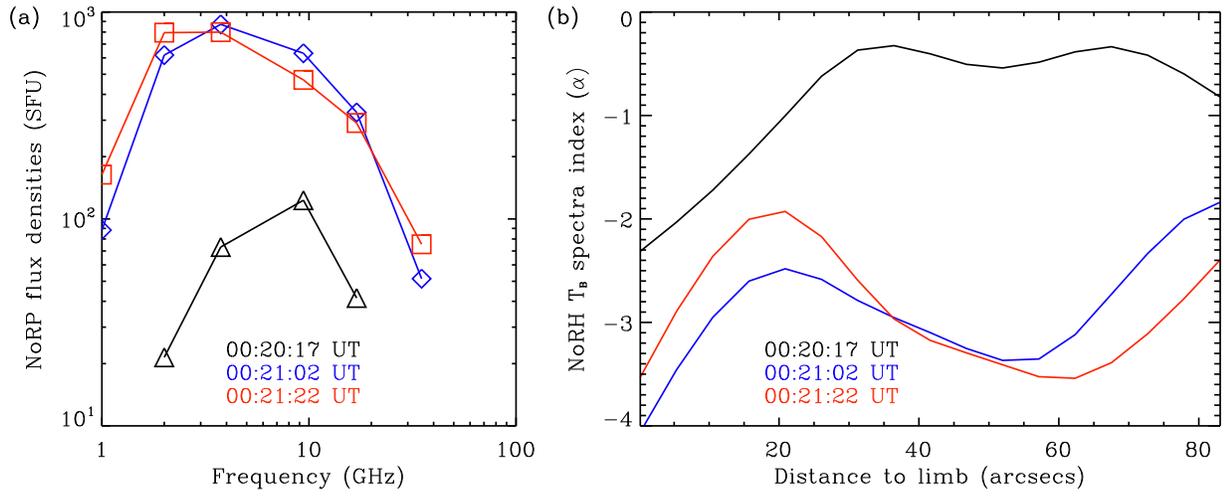}
\caption{(a) The NoRP flux densities from 1-35 GHz, and (b) the
NoRH 17-34 GHz spectral index ($\alpha$) of $T_B$ at three moments.}\label{Fig7}
\end{figure}

\end{document}